\documentclass[runningheads,fleqn]{cl2emult}

\usepackage{makeidx}  
\usepackage{graphicx} 
\usepackage{subeqnar} 
\usepackage{multicol} 
\usepackage{cropmark} 
\usepackage{phys}     
\makeindex            

\newcommand{\E}{{\mathrm e}}
\newcommand{\I}{{\mathrm i}}
\newcommand{\D}{{\mathrm d}}

\begin{document}
\title*{Influence of center-of-mass correlations on
spontaneous emission and Lamb shift in dense atomic gases$^1$}
%
%
%
%
\titlerunning{Spontaneous emission in dense atomic gases }
%
\author{M.~Fleischhauer\qquad
%
%
%
\institute{Sektion Physik,  
Universit\"at M\"unchen, 
D-80333 M\"unchen, Germany}
e-mail: mfleisch@theorie.physik.uni-muenchen.de}

\maketitle              

\begin{abstract}
Local field effects on  
the rate of spontaneous emission and Lamb shift in a dense gas 
of atoms are discussed taking into account correlations
of atomic center-of-mass coordinates.
For this the exact retarded propagator in the medium is calculated
in independent scattering approximation and employing a virtual-cavity
model. 
The resulting changes of the atomic polarizability 
lead to modifications of the
medium response which can be of the same order of magnitude 
but of opposite sign
than those due to local field corrections
of the dielectric function
derived by Morice, Castin, and Dalibard [Phys.Rev.A {\bf 51}, 3896 (1995)]. 
\footnote{This paper is dedicated to the memory of Dan Walls}
\end{abstract}


\section{Introduction}

The experimental progress 
in cooling and trapping of atoms and the observation 
of Bose-Einstein condensation \cite{BEC,BEC2} in atomic vapors
has lead to a growing theoretical interest in the
interaction of light with dense atomic gases 
\cite{opt_prop1,opt_prop2,opt_prop3,opt_prop4}. 
In the present paper it is analyzed how the spatial distribution of
nearest neighbors in a dense gas,
characterized by two-particle correlations, 
affects the interaction of an excited atom with the electromagnetic 
field vacuum. In particular modifications
of the rate of spontaneous emission and the Lamb 
shift are calculated and the resulting modifications of the
dielectric function discussed. 

While the interaction of light with a dilute gas is well described
in terms of macroscopic quantities 
with the well-known Maxwell-Bloch equations,
this is no longer true when the gas becomes dense. 
Here two new types of effects 
arise:
When the density of atoms
becomes large enough such that the resonant absorption length is less than
the characteristic medium dimension, re-absorption and
multiple scattering of spontaneous photons need to be taken into account. 
Secondly with increasing density dipole-dipole interactions 
between nearest neighbors become important. Here the macroscopic 
picture of a homogeneous polarization breaks down and it is necessary 
to introduce local field corrections.

The most famous local-field correction
leads to the Lorentz-Lorenz (LL) relation  between atomic 
polarizability $\alpha(\omega)$ and dielectric function $\varepsilon(\omega)$  
\cite{LL1,LL2}.
The LL-correction removes the
unphysical contact interaction of two atoms at the same position
which arises in a continuum picture of a homogeneous polarization
and is independent of any specifics of the atoms. Recently
Maurice, Castin, and Dalibard \cite{Maurice95} have derived a 
generalization of the LL relation that takes into account
center-of-mass correlations for finite distances 
at which point specific properties of the atomic gas enter. 
They showed in particular 
that the tendency of bosonic atoms to bunch,
when the critical temperature of condensation
is approached, leads to a measurable change of 
the complex refractive index. Even more pronounced
effects such as a dramatic line-narrowing were recently predicted by
Ruostekoski and Javanainen for a Fermi gas in the
case ideal case \cite{Ruostekoski99}
or in the presence of a BEC transition
\cite{Ruostekoski99b}.

On the other hand, it is known 
since the early work of Purcell \cite{Purcell46}, that the
microscopic environment of an excited atom can also change its interaction
with the field vacuum. Thus local field effects should
lead to a modification of the atomic polarizability itself,
in particular the spontaneous emission rate and the Lamb shift. Different 
macroscopic models for Lorentz-Lorenz-type corrections of 
the spontaneous emission rate of an atom
embedded in a {\it lossless dielectric} have been developed
\cite{Nienhuis76,Knoester89,Milonni95,Glauber91,deVries98} 
and experimentally tested \cite{Rikken95,Schuurmans98}. In the presence
of losses, as is the case in dense gases of the same kind of atoms,  
macroscopic \cite{Scheel99a,Scheel99b} and microscopic 
approaches \cite{Fleischhauer99b} have indicated however that
local-field effects due to
nearest neighbors may be equally important as the LL correction
of the contact interaction. 
In the present paper I analyze these effects
 using a Greens function approach.

According to Fermi's golden rule the rate of spontaneous 
emission $\Gamma$ and the Lamb shift
$\Delta$ are given by the
(regularized) exact retarded propagator ${\bf D}$ 
of the electric field 
at the position ${\vec r}_0$ of the probe atom 
\cite{Fleischhauer99b,Fleischhauer99a,Barnett92}. 
\begin{eqnarray}
\Gamma &=& \frac{2}{\hbar^2} \,
   {\vec d}\cdot{\rm Re}\Bigl[ {\bf D} 
      ({\vec r}_0,{\vec r}_0;\omega_0)\Bigr]\cdot{\vec d},\\
\Delta\,  &=& \frac{1}{\hbar^2} \,
  {\vec d}\cdot {\rm Im} \Bigl[ {\bf D} 
      ({\vec r}_0,{\vec r}_0;\omega_0)\Bigr]\cdot{\vec d}.
\end{eqnarray}
${\vec d}$ is the dipole vector and $\omega_0$ the 
true transition frequency of the atom. 
From the known propagator in free space and  after regularisation
one finds for an isolated atom
\begin{eqnarray}
\Gamma=\Gamma_0=\frac{d^2\omega_0^3}{3\pi\hbar\epsilon_0 c^3},
\qquad\qquad \Delta=0.
\end{eqnarray}
The exact retarded propagator in a medium can formally be obtained from a
scattering series. This series is here calculated 
neglecting multiple scattering of photons by the same atoms and 
taking into account only two-particle correlations
of center-of-mass coordinates.
It is shown that the alterations of the atomic polarizability
$\alpha$ due to local-field corrections of spontaneous emission and
Lamb shift lead to modifications of $\varepsilon$ which are of the same
order of magnitude as those found by
Maurice, Castin and Dalibard  \cite{Maurice95} and Ruostekoski and Javanainen 
\cite{Ruostekoski99,Ruostekoski99b}.


\section{Scattering series for the retarded propagator}


A dense medium affects the interaction of an excited probe atom with the
surrounding electromagnetic vacuum by multiple scattering
of  virtual photons emitted and re-absorbed 
by that atom. These scattering processes can formally be described by the
exact retarded Greens-function (GF) 
\begin{equation}
{\bf D}({\vec r}_1,{\vec r}_2 ;\tau)
=\theta(\tau)\bigl\langle 0 \bigr\vert 
\bigl[{\hat{\vec E}}(\vec r_1,t_1),
{\hat{\vec E}}(\vec r_2,t_2)\bigr] 
\bigl\vert 0\bigr\rangle,
\end{equation}
where $\tau=t_1-t_2$ and
${\hat{\vec E}}$
is the operator of the electric field interacting with all atoms. 
The free-space or vacuum GF in the frequency domain is given by
\cite{Pauli,deVries98b}
\begin{eqnarray}
{\bf G}^0(\vec x,\omega) &=&
-k^2\,
\frac{{\rm e}^{\I (k+\I 0) x}}{4\pi x}\left[
P\left(\I k x\right)\, {\bf 1} 
+Q\left(\I k x\right)\frac{{\vec x}\circ{\vec x}}{x^2}
\right] +\frac{1}{3} \delta(\vec x)\, {\bf 1},
\label{D0coord}
\end{eqnarray}
where ${\bf D}^0=\I\hbar {\bf G}^0/\epsilon_0$. 
Here $k=\omega/c$, $x=|\vec x|$ and
\begin{eqnarray}
P(z) = 1-\frac{1}{z} +\frac{1}{z^2},\qquad
Q(z) = -1 + \frac{3}{z} -\frac{3}{z^2}.\label{P-Q}
\end{eqnarray}
${\bf G}^0({\vec x},\omega)$ diverges as $x\to 0$ which is
related to the large-q behavior in reciprocal space. 
This will lead to corresponding divergences in the exact propagator
${\bf G}({\vec x},\omega)$ which can however be removed 
by introducing 
a regularisation ${\bf G}({\vec q},\omega)\to  {\bf G}({\vec q},\omega) 
\, f(\Lambda,q)$ with a wave-number cut-off $\Lambda$.
For the purpose of 
the present paper I will assume that the large-q behavior
of the Greens function is properly regularized and ignore all
contributions containing the regularisation parameter $\Lambda$.
The subject of regularisation will be discussed in more detail at a
different place.

The net effect of all possible 
multiple scattering events can be described by 
a scattering series for the exact propagator.  
We here assume not too large densities, such that
dependent scattering can be neglected. I.e. there can be
as many scattering events as possible but never twice from the same atom. 
In this so-called independent
scattering approximation (ISA), the scattering series 
can be expressed in the form
\begin{eqnarray}
&&{\bf G}(0,\omega) =
{\bf G}^0(0,\omega) -
\sum_{i\ne 0} {\bf G}^0({\vec x}_{0i},\omega)
\cdot{\bf a}_i(\omega) \cdot{\bf G}^0({\vec x}_{i0},\omega)
+\nonumber\\
&&\quad +\sum_{i\ne j\ne 0}
{\bf G}^0({\vec x}_{0i},\omega)\cdot{\bf a}_i(\omega)\cdot
{\bf G}^0({\vec x}_{ij},\omega)\cdot {\bf a}_j(\omega)\cdot 
{\bf G}^0({\vec x}_{j0},\omega) +\cdots,
\end{eqnarray}
where ${\bf a}_j(\omega)$ is the polarizability tensor of the $j$th atom
of the host material,
the summation  is over all atomic positions, and 
${\vec x}_{ij}={\vec r}_i-{\vec r}_j$.
It should be noted that the polarizability of the {\it excited}
probe atom does not enter the scattering series. 
However, the  probe atom can affect the
spatial distribution of the surrounding scatterers and
it is necessary to keep track of its presence.

We now assume a homogeneous medium of density $\varrho$
with randomly oriented two-level atoms 
such that $a_i(\omega)=\alpha(\omega)\, {\bf 1}$.
In this case we may replace the
sums over atomic positions 
by integrals. For this we introduce normalized joint
probabilities $p_2({\vec r}_0,{\vec r}_1)$, 
$p_3({\vec r}_0,{\vec r}_1,{\vec r}_2)$ etc.  
to find one atom at position ${\vec r}_1$ and
the  probe atom at ${\vec r}_0$; to find two
atoms  at positions ${\vec r}_1$ and ${\vec r}_2$
and the probe atom at ${\vec r}_0$ etc, with $p_1({\vec r})=1$.
This leads to
\begin{eqnarray}
&&{\bf G}(0) = 
{\bf G}^0(0) -\varrho{\alpha}\int \!\D^3{\vec r}_i\enspace
p_2({\vec r}_0,{\vec r}_i)\, {\bf G}^0({\vec x}_{0i})
\cdot{\bf G}^0({\vec x}_{i0})
+\nonumber\\
&& + \varrho^2{\alpha}^2\int\!\!\int\! \D^3{\vec r}_i\, \D^3{\vec r}_j\enspace
p_3({\vec r}_0,{\vec r}_i,{\vec r}_j)
\,{\bf G}^0({\vec x}_{0i})\cdot
{\bf G}^0({\vec x}_{ij})\cdot 
{\bf G}^0({\vec x}_{j0}) +\cdots\label{scatter}
\end{eqnarray}
where we have suppressed the frequency argument for notational simplicity.

For a dilute gas the positions of the atoms can be treated as independent
and one can factorize all particle correlations, which
amounts to $p_m\equiv 1$. 
The scattering series can then easily be solved.
The poles of ${\bf G}({\vec q},\omega)=
\int \!\D^3{\vec x}\, \E^{\I {\vec q}\cdot{\vec x}}\, {\bf G}({\vec x},\omega)$
determine the dielectric function for which one finds the well-known 
dilute-medium result
\begin{eqnarray}
\varepsilon(k)=\frac{q_0^2}{k^2}=1+\varrho\alpha(\omega).\qquad\qquad
(\omega= kc)
\end{eqnarray}
Properly regularizing ${\bf G}({\vec q},\omega)$ and
transforming the result back to coordinate space eventually
yields the spontaneous emission rate and the Lamb shift relative to
the vacuum
\begin{eqnarray}
\Gamma=\Gamma_0 \,{\rm Re}\,\bigl[\sqrt{\varepsilon}\bigr],
\qquad\Delta =
\frac{\Gamma_0}{2}\,{\rm Im}\,\bigl[\sqrt{\varepsilon}\bigr].
\end{eqnarray}
Here $\varepsilon$ is the dielectric function of the gas at the
true transition frequency.
For an atom embedded in a dielectric host with real dielectric function the 
result for $\Gamma$ is identical to that obtained by
Nienhuis and Alkemande based on a density-of-states argument 
\cite{Nienhuis76}.


\section{Local field effects and center-of-mass correlations}


In order to obtain a non-perturbative result for the retarded propagator
in a dense gas, one has to sum all contributions of the 
scattering series (\ref{scatter}) without factorizing the center-of-mass
correlations. Apart from some special cases such as hard-sphere scatterers
\cite{Lagendijk97} it is not possible to bring the scattering series
in an exact closed form and approximations are needed. 
The first approximation I use here is to take into
account only two-particle correlations by  using a Kirkwood-type
factorization \cite{Hansen76} of higher-order contributions
\begin{eqnarray}
p_3(1,2,3)=p_2(1,2)\, p_2(2,3)\, p_2(1,3),\quad {\rm etc.}
\end{eqnarray}
Furthermore it is assumed that only  
correlations between successive scatterers
matter, which is however correct up to second order in the density. 
With this the scattering series (\ref{scatter}) can be represented in
the form
\begin{eqnarray}
{\bf G}(0) &=& 
{\bf G}^0(0) -\varrho{\alpha}\, {\bf H}^0({\vec x}_{0i})
\cdot{\bf G}^0({\vec x}_{i0})+\nonumber\\
&&+\varrho^2{\alpha}^2 
\,{\bf H}^0({\vec x}_{0i})\cdot
{\bf H}^0({\vec x}_{ij})\cdot 
{\bf H}^0({\vec x}_{j0}) -\label{scatter2}\\
&&-\varrho^3\alpha^3\, {\bf H}^0({\vec x}_{0i})\cdot{\bf H}^0({\vec x}_{ij})\cdot {\bf H}^0({\vec x}_{jk})\cdot {\bf H}^0({\vec x}_{k0})+\cdots,
\nonumber
\end{eqnarray}
where the spatial integration has been suppressed and
\begin{eqnarray}
{\bf H}^0({\vec r}_1,{\vec r}_2,\omega) = 
p_2({\vec r}_1,{\vec r}_2)\,{\bf G}^0({\vec r}_1,{\vec r}_2,\omega)
\end{eqnarray}
is the retarded propagator modified by the two-particle correlation
$p_2$. Note that the first order term contains only a single function
${\bf H}^0$.

Since two-particle correlations between atoms of the host
material can be different from correlations 
between the (excited) probe atom and a host atom, 
we will distinguish these two in the following. 
This also includes the case of an atomic impurity in
an environment of a different species.
The scattering series can then be written in the form
\begin{eqnarray}
&&{\bf G}(0,\omega)={\bf G}^0(0,\omega)-
\varrho\alpha  \int\!\!\!\D^3{\vec r}_1
{\bf H}^0_{e}({\vec r}_0,{\vec r}_1,\omega)\cdot 
{\bf G}^0({\vec r}_1,{\vec r}_0,\omega)+
\nonumber\\
&&\quad+\int\!\!\!\int\!\!\D^3{\vec r}_1\,\D^3{\vec r}_2\,
{\bf H}^0_{e}({\vec r}_0,{\vec r}_1,\omega)\cdot 
{\bf T}^{(2)}({\vec r}_1,{\vec r}_2,\omega)
\cdot{\bf H}^0_{e}({\vec r}_2,{\vec r}_0,\omega),\label{HDyson}
\end{eqnarray}
where ${\bf T}^{(2)}({\vec r}_1,{\vec r}_2,\omega)$ is the 
part of the scattering matrix that contains at least two
scattering processes in the host material.
${\bf H}^0_e$ is the free propagator modified by the correlation
of the excited probe atom with an atom of the background gas. 
Although in ISA the ${\bf T}$-matrix does not contain scattering events from
the probe atom, it would in general still depend on its presence through 
the position correlations. With the earlier assumption that only 
correlations between successive scatterers matter, this dependence is lost.
 I.e. we treat the scattering of photons
in the gas as if the place of the probe atom would be filled
with the host material, which is equivalent to the 
{\it virtual cavity model} of Knoester and
Mukamel \cite{Knoester89}.
 From the discussion of impurities 
in cubic dielectric host materials by deVries
and Lagendijk \cite{deVries98} it is however 
expected that this approximation
does not affect the results in leading order of the density. 
Finally we will restrict ourselves  to the case of a
homogeneous and isotropic gas, such that the two-particle correlation 
depends only on the distance between the atoms
$p_2({\vec r}_1,{\vec r}_2)
=p_2(x_{12})$ where $x_{12}=|{\vec r}_1-{\vec r}_2|$.
In this case the scattering matrix obeys a simple Dyson equation in
reciprocal space
\begin{eqnarray}
{\bf T}({\vec q},\omega)=\varrho\alpha(\omega){\bf H}^0_{g}({\vec q},\omega)
\varrho\alpha(\omega) - \varrho\alpha(\omega){\bf H}^0_{g}({\vec q},\omega)
\cdot{\bf T}({\vec q},\omega)
\label{T}
\end{eqnarray}
${\bf H}^0_{g}$ is the free propagator modified by the
two-particle correlations of the host material.

It is convenient at this point to introduce the
irreducible correlation $h_2^\mu$ according to $p_2^\mu=1+h_2^\mu$. 
One then has
\begin{equation}
{\bf H}^0_\mu({\vec q},\omega) = -\frac{\bigl(\frac{1}{3} q^2 
+\frac{2}{3}k^2\bigr)\, {\bf 1}
-{\vec q}\circ{\vec q}}
{q^2-k^2-\I 0}+f_1^\mu(q,\omega)\, {\bf 1} +f_2^\mu(q,\omega) \, 
\frac{{\vec q}\circ{\vec q}}{q^2},\label{H0}
\end{equation}
with
\begin{eqnarray}
f_1^\mu(q,\omega) &=& -k^2\int_0^\infty\!\!\D x\, x \E^{\I k x}\, h_2^\mu(x)
\Bigl[j_0(qx) P(\I kx) + \frac{j_1(qx)}{qx} Q(\I kx)\Bigr],\\
f_2^\mu(q,\omega) &=& k^2\int_0^\infty\!\!\D x\, x \E^{\I k x}\, h_2^\mu(x)
\, j_2(qx)\, Q(\I kx).
\end{eqnarray}
$j_n(z)$ are spherical Bessel functions and $P$ and $Q$ have been defined
in eq.(\ref{P-Q}). 
In the first term of eq.(\ref{H0}) we have made use of 
$p_2^\mu({\vec x})\, \delta({\vec x})=0$ which corresponds to the
LL correction of the contact interaction.
The solution of eq. (\ref{T}) is now easy to obtain.
The poles $q_0$ of ${\bf T}$ which determine the dielectric function 
follow from the equation
\begin{equation}
q_0^2-k^2-\varrho 
\alpha(\omega)\left(\frac{1}{3}q_0^2+\frac{2}{3}k^2-f_1^g(q_0,\omega)
(q_0^2-k^2)\right)-\I 0=0.
\end{equation}
Since in lowest order of $\varrho\alpha$ one has
 $q_0^2\approx k^2+\I 0$, we may replace $f_1^g(q_0,\omega)$
by $f_1^g(k,\omega)$ which yields
\begin{equation}
\varepsilon(k)=\frac{q_0^2}{k^2}=1+\frac{\varrho\alpha(\omega)}
{1-\varrho\alpha(\omega)/3+\varrho\alpha(\omega)f_1^g(k,\omega)}.
\label{epsilon}
\end{equation}
This result  is identical to that 
of Maurice, Castin and Dalibard \cite{Maurice95} in ISA.
It is interesting to note that although the free GF ${\bf H}^0$
contained also longitudinal components (proportional to $f_2^g
{\vec q}\circ{\vec q}/
q^2$), they exactly cancel in the expression for the dielectric function.

To obtain the spontaneous emission rate and Lamb shift, we
consider only the leading order corrections in the density 
where we can  replace $f_{1,2}^\mu(q,\omega)$ by 
$f_{1,2}^\mu(k,\omega)$ and introduce a
regularisation of the large-q behavior. This yields
after some algebra
for the orientation averaged retarded GF:
\begin{eqnarray}
G({\vec x}=0,\omega)&=& -\frac{\I k^3}{6\pi}
\, \Biggl[1+\varrho\alpha\,\biggl(
\frac{7}{6} - f_1^e\biggr)+\nonumber\\
&&
+\varrho^2\alpha^2\, \biggl(\frac{17}{24}-\frac{7}{3} f_1^e
+{f_1^e}^2-\frac{7}{6} f_1^g 
+2f_1^ef_1^g\biggr)+\cdots\Biggr]
\end{eqnarray}
One recognizes that there is again no 
contribution from the longitudinal terms
$f_2$ up to second order in $\varrho$. 
Furthermore the two-particle correlation $f_1^g$ between ground-state
atoms enters only in second order of the density, while there
is a first-order contribution from $f_1^e$.  This is physically 
intuitive since correlations between ground-state atoms
enter only after two scattering events, while correlations
involving the probe atom are already important
in first order. 
From the above results one finds in leading order of the density
\begin{eqnarray}
\Gamma&=&\Gamma_0\,\biggl[1+\frac{7}{6}\varrho\alpha'
- \varrho (\alpha'{f_1^e}'-\alpha''{f_1^e}'')+\cdots\biggr],\label{Gamma_f}\\
\Delta&=&\frac{\Gamma_0}{2}
\,\biggl[\frac{7}{6}\varrho\alpha''- \varrho(\alpha''{f_1^e}'+
\alpha'{f_1^e}'')+\cdots\biggr]\label{Delta_f}
\end{eqnarray}
where we have introduced the real and imaginary parts of $f_1^e=
{f_1^e}'+\I {f_1^e}''$.
Eqs.(\ref{Gamma_f}) and (\ref{Delta_f}) are the main result of the
present paper. The first-order corrections to spontaneous emission
rate and Lamb shift, which are independent on the $f$'s are the
Lorentz-Lorenz local field corrections derived in
\cite{Fleischhauer99b}.

In order to illustrate the implications 
of eqs.(\ref{Gamma_f}) and (\ref{Delta_f}) to 
the medium response, I will discuss in the following section the
dielectric function of a dense gas of two-level atoms using some
simple model functions for the center-of-mass correlations.


\section{Modifications of medium response}


We have shown that local field corrections change not only
the relation between
atomic polarizability and dielectric function of the medium 
but also 
the atomic polarizability itself (with respect to the thin-medium case).
To illustrate the net effect of both corrections to the
dielectric function, I  now consider a classical, homogeneous gas of 
radiatively broadened (cold) two-level atoms
with randomly oriented dipole vectors. The dimensionless
atomic polarizability
of such atoms is isotropic and has in free space the strength 
\begin{eqnarray}
\overline{\alpha}\,\bigl(\overline{\delta}\, \bigr)=
\alpha\, k_0^3= \frac{6\pi}{\overline{\delta} -\I}
\end{eqnarray}
where $k_0$ is the resonance wavenumber, 
$\overline{\delta}=(\omega_{ab}-\omega)/\gamma_{ab}$ is the normalized detuning
from the {\it true} resonance 
and $\gamma_{ab}=\Gamma_0/2$ is the free-space dipole decay rate.
With this we find in lowest order of the dimensionless 
density $\bar\varrho=\varrho/k_0^3$ 
\begin{eqnarray}
\Gamma=\Gamma_0\,\biggl[1
+6\pi\,
\bar{\varrho}\,{f_1^e}''+\cdots\biggr],
\qquad\Delta=\frac{\Gamma_0}{2}
\,\biggl[7\pi\, \bar{\varrho}-6\pi\,\bar{\varrho}\, {f_1^e}'
+\cdots\biggr],\label{Delta_2NS}
\end{eqnarray}
Substituting these expressions into the dielectric function,
eq.(\ref{epsilon}),
 one eventually
finds for the shift relative to the dilute-medium resonance
and the effective linewidth up to first order in ${\bar\varrho}$:
\begin{eqnarray}
\gamma_{\rm eff}&=&\frac{\Gamma_0}{2}
\Bigl[1- 6\pi\, \bar{\varrho}\,{f_1^g}''+ 6\pi \bar{\varrho}\,{f_1^e}''
+\cdots\Bigr],\\
\Delta_{\rm eff} &=& \frac{\Gamma_0}{2}
\Bigl[-2\pi\,\bar{\varrho} + 6\pi\,\bar{\varrho}\, {f_1^g}'
 +7\pi\, \bar{\varrho}
- 6\pi\,\bar{\varrho}\, {f_1^e}'
+\cdots\Bigr]\label{Delta_eff}
\end{eqnarray}
The second term in the 
expression for the linewidth 
is due to local field corrections of the 
dielectric function and the third one due
to the changed spontaneous emission rate. 
It is interesting to note that both contributions
are of the same order of magnitude but differ in sign. Thus 
local field effects to the vacuum interaction may compensate
the line-narrowing / broadening effects 
resulting from  local field corrections
of the dielectric function in lowest order of the density.
In the expression for the line-shift, eq.(\ref{Delta_eff}), one recognizes the
familiar Lorentz-Lorenz shift $-2\pi\bar\rho$. The second 
 term emerges again from local field corrections of the 
dielectric function and the two last terms are due
to modifications of the Lamb shift.

To illustrate the effect of the center-of-mass correlations
let us consider a gas with repulsive interaction such that 
the two-particle correlation $h_2$ is close to $-1$ over 
an effective correlation distance $z$ and then approaches 
zero. As simple  model functions
we use a Gaussian and a hyper-Gaussian
\begin{eqnarray}
h_2^{(a)}(x)=-\exp\left[-\bar{x}^2/\bar{z}^2\right]\quad{\rm and}\quad
h_2^{(b)}(x)=-\exp\left[-\bar{x}^8/\bar{z}^8\right]
\end{eqnarray}
where the distance $\bar{x}= x/\lambda$ and 
the correlation length $\bar{z}=z/\lambda$ are normalized to the resonance 
wavelength $\lambda$.
I have plotted in the following figures the real and imaginary parts
of $f_1$ for both correlations. It is worth noting that for  
values of the correlation length 
larger than the resonance wavelength, 
$f'\rightarrow 7/12$, while $f''$ becomes a linear
function of $\bar{z}$. 
\begin{figure}
\includegraphics[width=.95\textwidth]{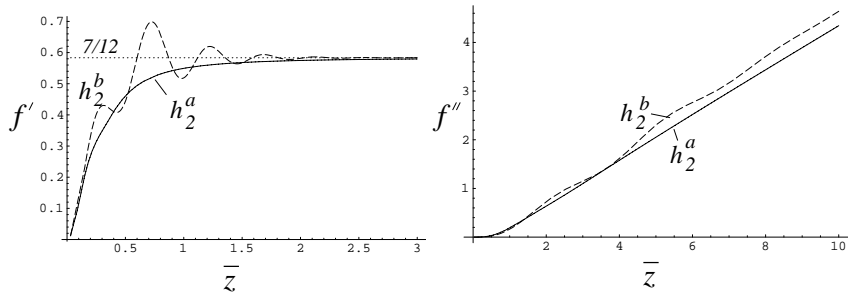}
\caption[width=.8\textwidth]
{Real ($f'$) and imaginary parts ($f''$) of $f$ 
as function of normalized correlation length
$\bar {z}=z/\lambda$ for Gaussian
and hyper-Gaussian correlation functions $h_2^{(a)}$ and
$h_2^{(b)}$}
\end{figure}

If the correlation length 
$\bar{z}$ scales with the density according to 
$\bar{z}\sim{\bar\rho}^{-1/3}$ 
the above behavior can be associated to the density dependence
of the correction terms.
In the following plots the real and imaginary parts of $\bar{z}^{-3} f_1
\sim \bar\varrho f_1$ are shown as function of $\bar{z}^{-3}\sim
\bar\varrho$.
\begin{figure}
\includegraphics[width=.95\textwidth]{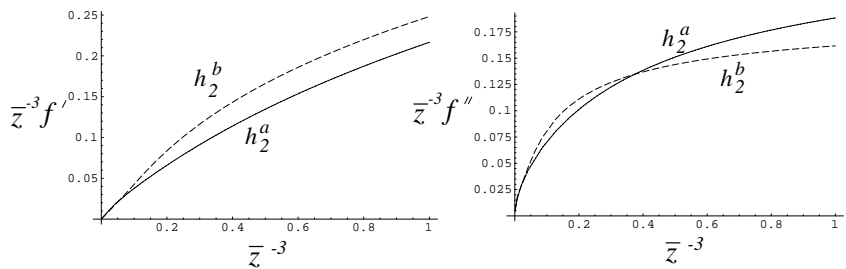}
\caption[width=.8\textwidth]
{Real and imaginary parts of ${\bar z}^{-3} f$ 
as function of 
$\bar {z}^{-3}\sim\bar\varrho$ for Gaussian
and hyper-Gaussian correlation functions $h_2^{(a)}$ and
$h_2^{(b)}$}
\end{figure}
One recognizes that  $\bar\varrho f_1''$
scales for small $\bar\varrho$ as $\bar\varrho^{2/3}$, while
$\bar\varrho f_1'$ is approximately linear in the density.
Thus in the
low-density limit linewidth changes will dominate line shifts.


\section{summary}


In the present paper local field effects on spontaneous emission and
Lamb shift in a dense atomic gas have been discussed taking into account
two-particle center-of-mass correlations. It has been shown that the
corresponding changes of the atomic polarizability can lead to
modifications of the medium response, which are of the same order as 
those resulting from direct corrections of the dielectric function
found by Maurice, Castin and Dalibard \cite{Maurice95}
and Ruostekoski and Javanainen \cite{Ruostekoski99}.
They are however of opposite sign
and thus may compensate the leading order corrections in the density. 
The present approach is based on an independent scattering approximation 
and a virtual cavity assumption. 
It is thus only applicable for densities which are sufficiently 
smaller than the cubic
wavenumber, $\bar\varrho\ll 1$.
Also the existence of a distinguished probe atom has been assumed, which
is valid only for classical gases. Nevertheless the results indicate
that local field corrections to the atomic polarizability
may change or even reverse the predicted line-shifts and
linewidth modifications found for Bose gases near the
condensation temperature and for low-temperature Fermi gases.


\clearpage
\addcontentsline{toc}{section}{Index}
\flushbottom
\printindex

\end{document}